\documentclass[a4paper,fleqn,usenatbib]{mnras}

\usepackage{newtxtext,newtxmath}

\usepackage[T1]{fontenc}
\usepackage{ae,aecompl}

\usepackage{graphicx}	
\usepackage{amsmath}	
\usepackage{amssymb}	

\newcommand{\dechms}[4]{$#1^{\rm h}#2^{\rm m}#3\mbox{$^{\rm s}\mskip-7.6mu.\,$}#4$}

\newcommand{\decdms}[4]{$-#1^{\circ}#2'#3\mbox{$''\mskip-7.6mu.\,$}#4$}

\title[G5.89:  An Explosive Outflow Powered by a Proto-Stellar Merger?]{G5.89:  An Explosive Outflow Powered by a Proto-Stellar Merger?}
\author[Zapata et al.]{
Luis A. Zapata,$^{1}$\thanks{E-mail: l.zapata@irya.unam.mx}
Paul T. P. Ho,$^{2,3}$
Estrella Guzm\'an Ccolque,$^{4}$
Manuel Fern\'andez-Lop\'ez,$^{4}$ \newauthor
Luis F. Rodr\'\i guez,$^{1,6}$  
John Bally,$^{5}$
Patricio Sanhueza,$^{7}$
Aina Palau,$^{1}$
and Masao Saito$^{7}$\\ \\
$^{1}$Instituto de Radioastronom\'\i a y Astrof\'\i sica, Universidad Nacional Aut\'onoma de M\'exico, P.O. Box 3-72, 58090, Morelia, Michoac\'an, M\'exico\\
$^{2}$Academia Sinica Institute of Astronomy and Astrophysics, PO Box 23-141, Taipei, 10617, Taiwan\\
$^{3}$East Asian Observatory, 666 N. A'ohoku Place, Hilo, Hawaii 96720, USA\\
$^{4}$Instituto Argentino de Radioastronom\'\i a (CCT-La Plata, CONICET; CICPBA), C.C. No. 5, 1894, Villa Elisa, Buenos Aires, Argentina\\
$^{5}$Astrophysical and Planetary Sciences Department University of Colorado, UCB 389 Boulder, Colorado 80309, USA\\
$^{6}$Mesoamerican Centre for Theoretical Physics, Universidad Aut\'onoma de Chiapas, Carretera Emiliano Zapata Km. 4\\ 
          \noindent Real del Bosque, 29050 Tuxtla Guti\'errez, Chiapas, M\'exico\\
$^{7}$National Astronomical Observatory of Japan, National Institutes of Natural Sciences, 2-21-1 Osawa, Mitaka, Tokyo 181-8588, Japan\\
}

\date{Accepted XXX. Received YYY; in original form ZZZ}

\pubyear{2015}

\begin{document}
\label{firstpage}
\pagerange{\pageref{firstpage}--\pageref{lastpage}}
\maketitle

\begin{abstract}
The explosive outflows are a newly-discovered family of molecular outflows associated with high-mass star forming regions. Such energetic events are  
possibly powered by the release of gravitational energy related with the formation of a (proto)stellar merger or a close stellar binary.  Here, we present 
sensitive and high angular resolution observations (0.85$''$) archival CO(J=3-2) observations carried out with the Submillimeter Array (SMA) 
of the high-mass star forming region G5.89$-$0.39 that 
reveal the possible presence of an explosive outflow.  We find six well-defined and narrow straight filament-like ejections pointing back approximately 
to the center of an expanding molecular and ionized shell located at the center of this region. These high velocity ($-$120 to $+$100 km s$^{-1}$) 
filaments follow a Hubble-like velocity law with the radial velocities increasing with the projected distance. The estimated kinematical age 
of the filaments is about of 1000 yrs, a value similar to the dynamical age found for the expanding ionized shell.   G5.89 is the
thus the third explosive outflow reported in the galaxy (together with Orion BN-KL and DR21) and argues in favor of the idea that this is a frequent 
phenomenon. In particular, explosive outflows, in conjunction with runaway stars, demonstrate that dynamical interactions in such groups 
are a very important ingredient in star formation.

\end{abstract}




\begin{figure*}
	\includegraphics[scale=0.56]{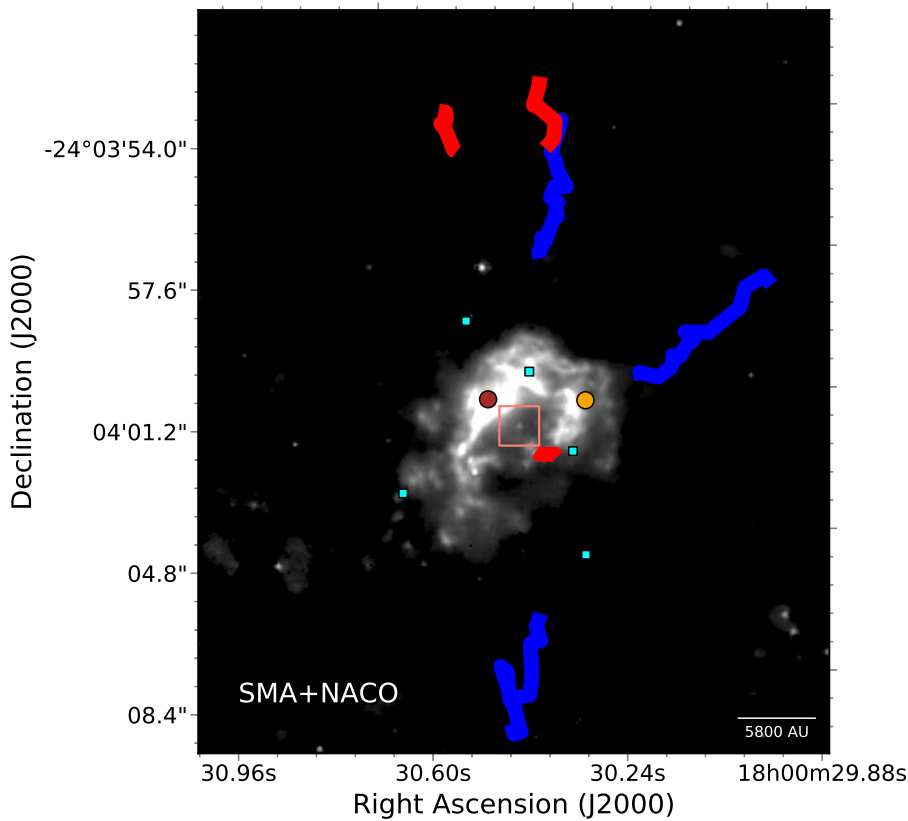}
    \caption{Approaching (blue) and receding (red) molecular filaments in G5.89 as observed with the SMA, overlaid on a NACO/VLT "L" band infrared image
                  \citep{fel2003}. As in Orion-KL or DR 21, each colored line or filament represents series of positions at which CO emission peaks in consecutive velocity
                  spectral channels. The CO peaks in every channel can be seen in the Appendix.  All filaments point to the position $\alpha_{J2000.0}$ = \dechms{18}{00}{30}{3} $\pm$ 1.0$''$, 
                  $\delta_{J2000.0}$ = \decdms{24}{04}{01}{5} $\pm$ 1.0$''$. This central position is marked with a pink square and corresponds to the center of the infrared shell. 
                  The position of the objects reported in \citet[][]{fel2003} and \citet[][]{pug2006} are marked with a red (Feldt's star) and yellow (Puga's object) circle, respectively.
                  The cyan squares mark the position of the SMA submillimeter compact sources reported by \citet{hun2008}. } 
    \label{Fig1}
\end{figure*}

\section{Introduction}

G5.89$-$0.39 or W28 A2 (hereafter G5.89) is an expanding shell-like Ultracompact HII (UC HII) region \citep{woo1989} at a distance of 2.9$^{+0.19}_{-0.17}$ kpc \citep{sat2014}. 
This measure is inconsistent from that obtained in \citet[][]{moto2011} using similar techniques (1.28$^{+0.09}_{-0.08}$ kpc), however, the value obtained by \citet{sat2014}
seems more reliable since they used two different background sources limiting the potential position errors of the masers as is described in their study. 
The UC HII region is about 0.1 pc (7$''$) in size, and its dynamical age is 600$^{+250}_{-125}$ years, estimated from the angular expansion rate \citep{aco1998}. 
Combining a model for the nebular emission and H radio recombination line spectra, \cite{aco1998} estimated an expansion velocity for the UCHII region 
of about 35 km s$^{-1}$. Near infrared NACO-VLT observations carried out by \citet{fel2003} revealed that G5.89 contains a young O5 V star, which they proposed
to be the exciting source of the UC HII region. This young star, however, is located about 1$''$ to the northwest from the center of the shell-like UC HII region. 
\citet{pug2006} also reported an ionized outflow traced by the Br$\gamma$ emission located in the northeast side of the UC HII region. This outflow is likely tracing the 
presence of a second object associated with the region. There are also a group of (sub)millimeter compact sources located in the vicinities of the 
UC HII region \citep{sol2004,su2012}. However, as it is discussed in \citet{hun2008}, some of them could not be heated internally.   

It is also interesting to note that there are also many X-ray energetic sources, far infrared and a GeV to gamma-ray 
source (HESSJ1800-240B) associated with G5.89 \citep{ham2016,gus2015,leu2015}. In particular, far-IR spectroscopic Herschel observations obtained
by \citet{kar2014} revealed high-J CO lines where most of the CO luminosity from these outflows is possibly radiated.

G5.89 is associated with high-velocity ($-$75 to $+75$ km s$^{-1}$)  outflowing gas originally identified in CO and CS by \citet{har1988}. 
The outflow emission has been subsequently studied using single-dish CO and SiO \citep{kla2006} and interferometric
CO, HCO$^+$, and SiO \citep{wat2007,sol2004} observations. \citet{kla2006} proposed that
the outflow located in this region is possibly a fossil flow (that is, a flow without a present excitation source but that continues moving by momentum conservation) 
with an age of about 1000 yrs.  The reported position angle of the outflow is notably different between the tracers, being nearly east-west in CO and HCO$+$
(position angle $+$84$^\circ$), vs. northeast-southwest in SiO (position angle $+$30$^\circ$). 
The origin and relation of these different outflows remain poorly understood. Moreover, the powerful SiO bipolar outflow
reported by \citet{sol2004} may be a remanent outflow ejected possibly from Feldt's infrared star or some other 
protostars in the vicinity of the UCHII region. This is because recent SMA high angular observations resolved both lobes \citep[see Figure 8 from][]{hun2008}. 
On the other hand, the outflows reported in this region are massive and very energetic. For example, \citet{har1988} reported 
that the outflow located in this region is about 1000 times more energetic ($\sim$10$^{49}$ erg) than the one located 
in Orion-KL ($\sim$10$^{46}$ erg).  For the case of the outflow in Orion-KL, \citet{goi2015} reported that the L(CO)/L$_{bol}$ luminosity ratio in this region
is quite different compared to other high-mass SFRs, making the Orion-KL outflow peculiar. 

More recently, using interferometric SMA sensitive observations \citet{su2012} reported extremely high-velocity  ($-$150 to $+80$ km s$^{-1}$) 
outflowing gas in CO (J=2-1) and (J=3-2) associated with the shell-like UC H II region. 
These high velocity lobes were interpreted as belonging to two different outflows. 
In addition, the outflow lobes clearly show a Hubble-like kinematic structure. A diagram of the estimated temperature 
of the outflowing gas vs. their radial velocity of the molecular lobes presented in Figure 4 of \citet{su2012} showed that the
temperature increases with the radial velocities and projected distances. This result suggests that the outflowing gas is
warmer on the tips possibly as a result of strong shocks with the interstellar medium.   

The explosive outflows are a new family of molecular outflows associated with high-mass star forming regions \citep{zap2009,zap2013,ball2016,ball2017}. 
Such energetic events are possibly powered by the release of gravitational energy related with the formation of a 
(proto)stellar merger or a close stellar binary \citep{ball2005,zap2017,ball2017}.
At this moment, there are four clear morphological and kinematical differences between the classic protostellar and 
the newly-discovered explosive molecular outflows that are described in \citet{zap2017}:

\begin{itemize}
\item The explosive outflows consist of narrow straight filament-like ejections or streamers with different directions and in almost an isotropic configuration.
\item The narrow molecular filaments point back to approximately an explosion site.
\item The outflow presents a very-well-defined Hubble flow-like increase of velocity with distance from the origin in the explosive filaments.
\item The overlapping of the redshifted components with respect to the blueshifted components.
\end{itemize}   

In this paper, using archival SMA high angular resolution ($\sim$0.8$''$) CO(3-2) observations and following the
characteristics of the explosive outflows described in \citet{zap2017}, we propose the possible presence of an explosive outflow in G5.89
centered in the expanding ionized and molecular shell-like structure found in this region.  In the next sections, we describe in more detail our findings
on this remarkable region. 

\section{Archival Observations}

The SMA\footnote{The Submillimeter Array is a joint project between the Smithsonian Astrophysical Observatory and the Academia Sinica Institute 
of Astronomy and Astrophysics and is funded by the Smithsonian Institution and the Academia Sinica.} observations were carried out in 2006 
June 05 using its extended configuration.  At that time, the array was with its eight antennas. The 28 independent baselines in this configuration 
ranged in projected length from 15 to 233 k$\lambda$.
The phase center of the observations were situated at the position $\alpha_{J2000.0}$ = \dechms{18}{00}{30}{2}, $\delta_{J2000.0}$ = \decdms{24}{04}{00}{5}.
The SMA primary beam is approximately 30$''$ at 345 GHz. This allowed us to study the entire outflow in G5.89 with a single pointing. 
In these observations, we did not merge the SMA data with short-spacing observations (single-dish) because we were interested in the high-velocity compact emission
from the CO, and not with the extended systemic emission.

The digital correlator was set to have 24 spectral "chunks" of 104 MHz and 128 channels each. This spectral resolution
yielded a velocity bin width of about 0.7 km s$^{-1}$. However, we smoothed this to 7 km s$^{-1}$ given the very broad CO line present
in G5.89, see \citet{su2012}.  The observations were centered at a frequency of  346.48 GHz in the upper side band, while in the lower
side band was at 336.48 GHz. The CO(3-2) line was detected at a rest frequency of 345.81 GHz. The  total  available  double-sideband
bandwidth was 4 GHz.   

The quasars J1733$-$130 and J1911$-$201 were used as gain calibrators, while Jupiter's moon Callisto and the dwarf planet Ceres
were used as bandpass and flux calibrator, respectively. The uncertainty in the flux scale is estimated to be between 15\% and 20\%, 
based on the SMA monitoring of quasars.  

The data\footnote{The raw data can be obtained from: http://www.cfa.harvard.edu/} were calibrated  using the IDL superset MIR 
adapted for the SMA\footnote{The MIR-IDL cookbook by C. Qi can be found at http://cfa-www.harvard.edu/~cqi/mircook.html.}.   
The calibrated data were then imaged and analyzed in standard manner using $MIRIAD$ \citep{sau1995} and $CASApy$ \citep{mac2007}.
We also used some routines in Python to image the data \citep{ast2013}. More technical specifications of the SMA and its calibration strategies 
can be found in \citet{ho2004}. 

 A $^{12}$CO(J=3-2) spectral velocity cube was obtained setting the {ROBUST} 
parameter of the task {INVERT} to $+$2 to obtain a better sensitivity. 
The contribution from the strong continuum was subtracted using the MIRIAD task {\it uvlin}. 
The resulting r.m.s.\ noise for the line cube was about  30 mJy beam$^{-1}$ per velocity channel, 
with a beam with an angular resolution of $0\rlap.{''}9$ $\times$ $0\rlap.{''}8$ with a P.A. = $-10^\circ$.
In this paper, we concentrate in the CO(J=3-2) line emission from G5.89, the continuum and some 
other lines present in the observations (e.g. SO, SO$_2$ and H$_2$CO) are mostly associated 
with the ionized shell. We refer to the author to \citet{hun2008}, where a very dedicated SMA study is
presented of these structures. 
  
\section{Results}

The CO filament-like structures revealed in our Figures \ref{Fig1} and \ref{Fig2} were obtained from the maps in velocity windows
of  7 km s$^{-1}$ width presented in the Appendix. In these velocity channels maps we found about one hundred localized emission
features. The position and radial velocities of these condensations were then obtained using linearized least-squares fits to Gaussians ellipsoids
using the task SAD of AIPS.  This is a similar procedure to that used by \citet{zap2009,zap2013,zap2013b}. We discerned about six molecular
filaments that show consistent velocity increments and they are presented in Figures \ref{Fig1} and \ref{Fig2}.  It is important to mention that
we do not detected more molecular structures other than the filaments-like objects.

In Figure \ref{Fig1}, we present the most prominent high-velocity CO(J=3-2) features outside of the velocity window from $-$50 to $+$40 km s$^{-1}$,
overlaid on  a NACO/VLT "L" band infrared image \citep{fel2003} of G5.89. Within this velocity window  
the radiation arises basically from the ambient cloud (and probably some other molecular clouds close to the galactic plane given the broad range of velocities) 
and is spatially extended, thus cannot be completely reconstructed by The Submillimeter Array. The infrared image is tracing a shell-like structure
that is associated with the expanding UCHII region at the center of G5.89. Additionally, we also include in this image the positions 
of the two objects located in the vicinities of the infrared shell, the Feldt and Puga objects reported in
 \citet[][]{fel2003} and \citet[][]{pug2006}, respectively.  The compact infrared source in the middle of the rectangle 
is probably part of the shell, the H/K$_s$/L' image from \citet[][]{fel2003} does not reveal any massive young object. Here, the L' (3.8 $\mu m$) 
is tracing emission from warm dust from the shell and the possibly from the disk of the Feldt's star.
 
 Receding CO emission shows radial velocities up to $+$90 km s$^{-1}$,
 while approaching (blueshifted) radial velocities reached $-$110 km s$^{-1}$. 
 This is in very good agreement with the SMA observations of \citet{su2012}. 
 However, in our observations with better angular resolution 
 (a factor of almost 3 better than the previous SMA observations) all the CO 
 condensations were much better resolved angularly into narrow filament-like structures (Figure \ref{Fig1}).  Every CO 
 gas condensation reported in Figure 2 of \citet{su2012} has a counterpart in the form of a filament-like structure. The most northern
 redshifted condensation was resolved into two filaments. In total, we report the presence of six straight and narrow filaments, three redshifted and
 three blueshifted pointing approximately to the center of the infrared shell at the position 
 $\alpha_{J2000.0}$ = \dechms{18}{00}{30}{3} $\pm$ 1.0$''$ and $\delta_{J2000.0}$ = \decdms{24}{04}{01}{5} $\pm$ 1.0$''$.
 This position corresponds approximately to the center of the infrared and radio shell reported in \citet{fel2003,aco1998}, see Figure \ref{Fig1}. 
 We fitted very carefully straight lines to every filament and extrapolate them to find this position in the sky. Note  that none of 
 these molecular filaments points directly to the infrared and optical sources located in the surroundings of the UC HII region. 
    
A projected distance vs. radial velocity diagram of the six straight filaments reported in this study is presented in Figure \ref{Fig2}.
This diagram reveals the kinematics of the molecular filaments. Each filament follows a Hubble velocity law, that is, the radial velocity 
along every filament changes linearly with on-the-sky distance from this center. Moreover, all filaments seem to converge to the 
single radial velocity of about 9 km s$^{-1}$.  This velocity is approximately the systemic velocity of the cloud in G5.89, see \citet{su2012}.
 
 \begin{figure}
  	\centering
	\includegraphics[scale=0.51]{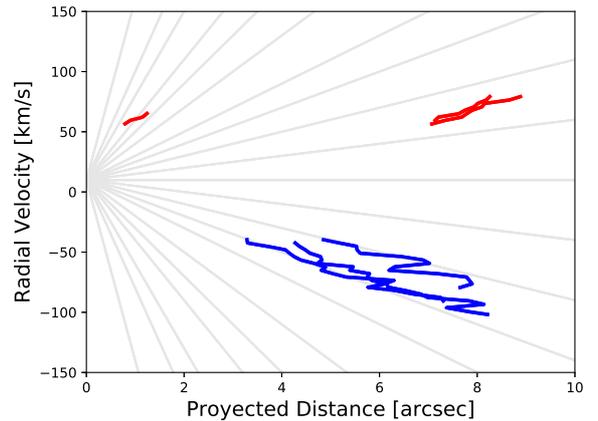}
    \caption{Radial velocity vs. projected distance diagram of the molecular filaments detected in G5.89. Blue lines trace the approaching filaments, while red lines are
                   tracing the receding ones.  Each colored line or filament represents series of positions at which CO emission peaks in consecutive velocity
                  spectral channels, see the channel maps (Figures \ref{fig3} and \ref{fig4}) in the Appendix. Gray gradient lines start from $+$9 km s$^{-1}$ at R$=$0. Note that velocities between 
                  $-$50 and $+$40 km s$^{-1}$ could not be studied because of interferometric contamination with extended molecular gas.} 
    \label{Fig2}
\end{figure}
 
\section{Discussion}

 All physical and kinematical features found on this outflow are reminiscent of those observed in two previously known explosive outflows: Orion-KL
 and DR21, see \citet{zap2017,ball2016}.  The flow found in G5.89 consists of narrow straight filament-like ejections with different directions. 
The six narrow molecular filaments point back to approximately the same central position. This central position is devoid of known sources at different wavelengths (IR, mm, cm...),
see our Figure \ref{Fig1}.
These filament-like structures present very-well-defined Hubble flow-like increase of velocity with distance from the origin, which is the center of the UCHII region.
There is overlapping in the northern region between the redshifted and the blueshifted components \citep{hun2008, su2012}. 
This overlapping can be even better discerned on the single-dish images presented in \citet{cho1993}.
We therefore propose that the outflow from G5.89 is explosive. 

The outflow in G5.89 however is far from being isotropic, as for example, the outflow in Orion-KL \citep{ball2017}.  This can be explained in terms  
of sensitivity.  SMA observations of the explosive outflow in Orion-KL revealed about 40 expanding molecular filaments associated with this flow \citep{zap2009}, 
and recent ALMA observations confirmed its explosive nature with more than 150 molecular filaments \citep{ball2017}. This difference in the number of filaments
mainly lies in the tremendous sensitivity offered by the ALMA telescope, approximately a factor of about 100 improvement in line emission sensitivity as compared to the SMA. 
ALMA observations of G5.89 could confirm the explosive nature of the flow, and define its overall geometry. 

The presence of an explosive outflow might explain some of the kinematical and morphological characteristics of G5.89.  For example,
the fact that this outflow is considered as a fossil outflow \citep{kla2006} is in agreement with a single and brief event that produced the molecular debris. 
The ionized and infrared expanding shell could also be created by this powerful event \citep[a molecular and expanding shell structure is also associated with the explosive outflow
in Orion-KL][]{zap2011}.  Moreover, the center of the ionized UCHII coincides with origin of the explosive flow, see Figure \ref{Fig1} and \citet{hun2008},
this indicated that the UC HII region could be a result of the explosion. A UCHII region at the center of the explosive outflow is not found in Orion-KL possibly
because this still very embedded in large quantities of material. But, for the case of DR21, there is the presence of a large HII region \citep{zap2013} with a similar age 
to that of the explosion. As this ionized region in DR21 is older (10$^4$ years), this is not very embedded in dust and one can see the ionized emission escaping from this region.   

We thus consider that the explosive outflow in G5.89 was possibly caused by a (proto)stellar merger or the formation of a close binary as in Orion-KL \citep{ball2016, ball2017}. We speculate  
that perhaps the O5 V young star, the Feldt's infrared star might be involved in this kind of energetic event.  Taking a mean radial velocity of 80 km s$^{-1}$ for the outflow 
at a distance from the origin of 6$''$ (this is a crude value for the length of the filaments), we found a kinematical age for the molecular flow of 1000 yrs, a value similar to the dynamical age found in 
the expanding ionized shell (600 yrs) and the fossil CO flow (1000 yrs) reported by \citet{kla2006}. Assuming a dynamical age of the explosion 
of 1000 yrs, and the distance from Feldt's star to the center of the explosion is about 1.5$''$ or 4300 AU, we estimated that the velocity in the plane of the sky
of this infrared star could be around 20 km s$^{-1}$. This is in very good agreement with the tangential velocities also in the plane of the sky of the runaway Orion stars BN, Source I,
and n, see \citet{rod2017}.  As the Feldt's star putative tangential velocity (20 km s$^{-1}$) coincides with the expanding velocity of the infrared 
and ionized UCHII region (35 km s$^{-1}$), we suggest that both objects were maybe ejected at a similar time.  In this picture, both objects should 
coincide with a similar timescale.  As the Puga's object is also at a similar 
distance from the explosion center, it is suggestive that maybe this object is also related to the outflow and the UCHII region. However, we think
that many more observations are needed to confirm this hypothesis.  

Even when future ALMA observations could confirm our findings in G5.89, the fact that we have a third explosive outflow in the galaxy argues in favor of 
       the idea that this is a common phenomenon happening in high-mass star forming regions.  With more cases like this one discovered in different SFR will allow us to
       estimate more precisely the rate of massive protostellar merging, something crucial on the formation of the massive stars as is suggested
       by the closest high-mass SFR: Orion-KL. 
       
In particular, explosive outflows, in conjunction with runaway stars, demonstrate that dynamical interactions in such groups are a very important ingredient in star formation.   
Interactions and dynamics can limit accretion and therefore set the final masses of stars.  N-body dynamics may be even more important in establishing the Initial Mass 
Function (IMF) than initial conditions.    The greater the number of explosive outflows, the more important N-body dynamic is for the IMF.     

We undertook an extensive search in the GAIA DR2 catalogues for the proper motions of the Feldt's star. However, at this point there seems to be no 
optical counterpart for this object. Deep radio observations with the VLA/SMA have also not revealed a counterpart in these wavelengths for this object \citep{hun2008}.        

\section{Conclusions}

In this paper, we have presented sensitive CO(J=3-2) archival observations from the high-mass star forming region G5.89 carried out 
with the Submillimeter Array (SMA). Given the good sensitivities and the sub-arcsecond angular resolution of these observations, we revealed six well-defined and narrow 
straight filament-like ejections pointing back approximately to the center of an expanding molecular and ionized shell located at the center 
of this region. These high velocity ($-$120 to $+$100 km s$^{-1}$) filaments follow a Hubble-like velocity law with the radial velocities increasing
 with the projected distance. These structures and kinematics have been reported to be present in explosive outflows as Orion-KL and DR21.
We conclude that the outflow in G5.89 is explosive and could be originated by a (proto)stellar merger or maybe the formation of 
a close binary, where perhaps the O5 V Feldt's star might be involved.

\section*{Acknowledgements}
LAZ and LFR acknowledge financial support from DGAPA, UNAM, and CONACyT, M\'exico.
A.P. acknowledges financial support from UNAM-PAPIIT IN113119 grant, M\'exico.
This research has made use of the SIMBAD database, operated at CDS, Strasbourg, France.
We are very grateful to Markus Feldt for having provided the NACO images.
We are really thankful for the thoughtful suggestions of the anonymous referee that 
helped to improve our manuscript.








\appendix

\section{Channel Maps}

Channel velocity maps of the CO(3-2) line emission obtained with the SMA from G5.89. 

\begin{figure}
	\includegraphics[scale=0.29]{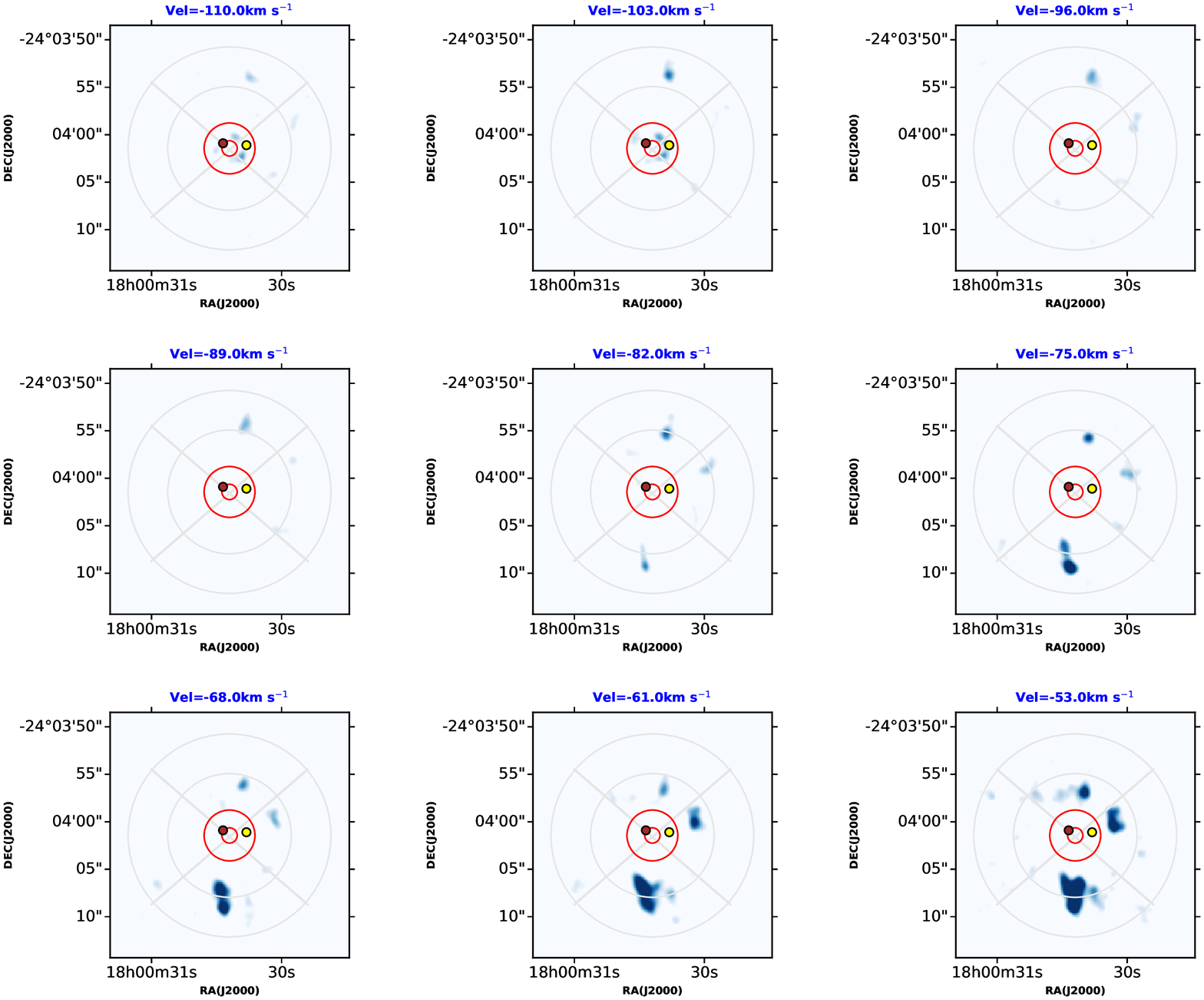}
    \caption{ SMA CO(J=3-2) channel velocity maps of the blueshifted emission from the outflow in G5.89. The radial velocities are indicated in the upper part 
    of each channel. The position of the objects reported in \citet[][]{fel2003} and \citet[][]{pug2006} are marked with a red (Feldt's star) and yellow (Puga's object) circle, respectively.
    The red rings in the center of each channel traces the position of the UCHII region, see \citet{aco1998}. The emission arising from the UCHII region at high velocities is due likely 
    to line contamination from some other molecular species. At the corresponding rest frequency 345.929 GHz (between $-$110 and $-$103 km s$^{-1}$), we identified the molecular line transition $^{34}$SO$_2$(17$_{4,14}$,17$_{3,15}$).  The synthesized beam size of the image is $0\rlap.{''}9$ $\times$ $0\rlap.{''}8$ with a P.A. = $-10^\circ$.
    The grey rings are only intended to guide the author in the position of the molecular CO condensations at different spectral channels.  } 
    \label{fig3}
\end{figure}
 
\begin{figure}
	\includegraphics[scale=0.27]{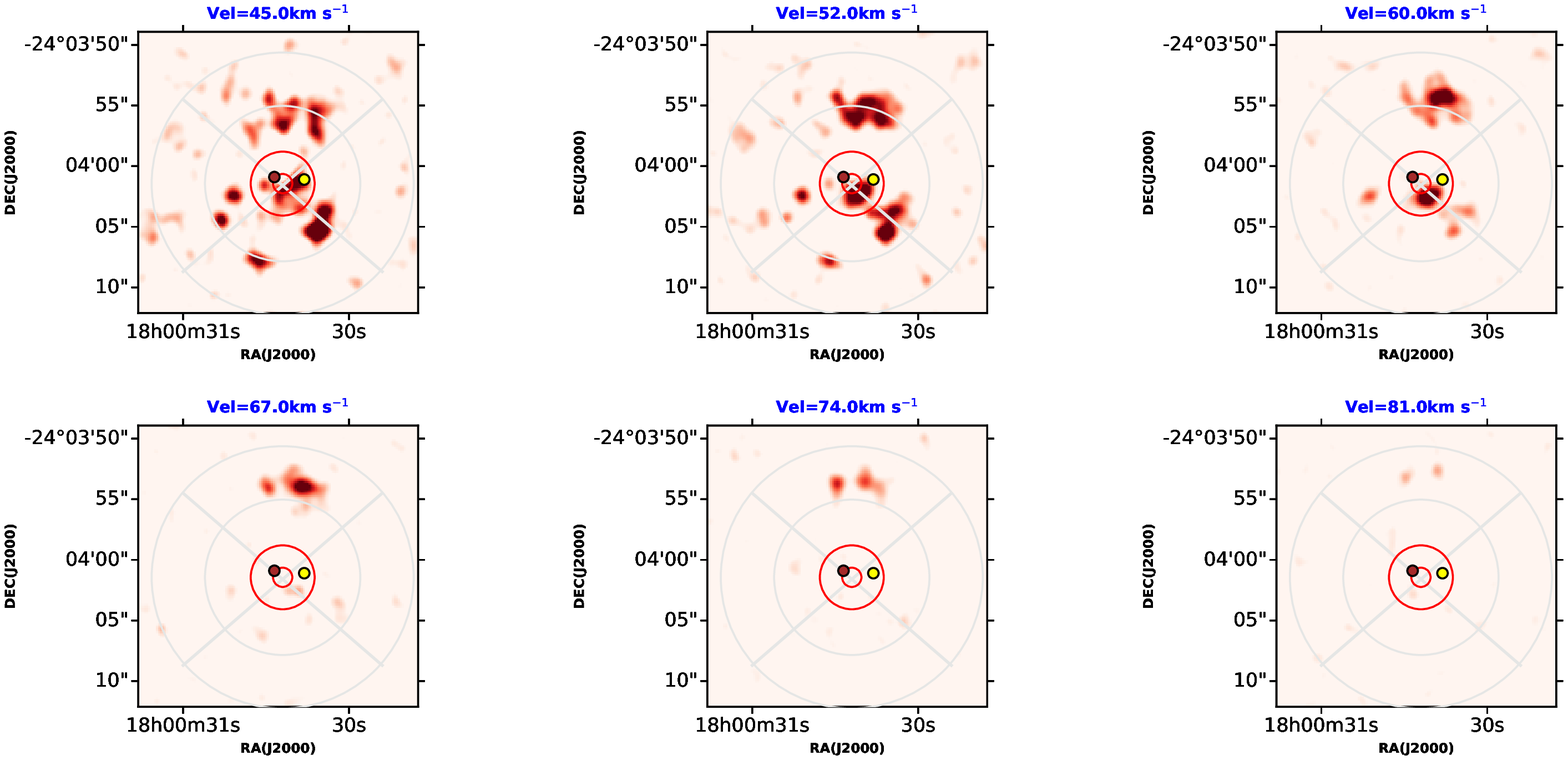}
    \caption{SMA CO(J=3-2) channel velocity maps of the redshifted emission from the outflow in G5.89. The radial velocities are indicated in the upper part
    of each channel. The position of the objects reported in \citet[][]{fel2003} and \citet[][]{pug2006} are marked with a red (Feldt's star) and yellow (Puga's object) circle, respectively.
    The red rings in the center of each channel traces the position of the UCHII region, see \citet{aco1998}. The synthesized beam size of the image is $0\rlap.{''}9$ $\times$ $0\rlap.{''}8$ with a P.A. = $-10^\circ$. The grey rings are only intended to guide the author in the position of the molecular CO condensations at different spectral channels. } 
    \label{fig4}
\end{figure}


\bsp	
\label{lastpage}
\end{document}